\documentstyle[prl,aps,twocolumn,epsf,floats]{revtex}
\begin{document}
\draft
\twocolumn[\hsize\textwidth\columnwidth\hsize\csname @twocolumnfalse\endcsname
\title{Quantum behavior of Orbitals in Ferromagnetic Titanates: \\
Novel Orderings and Excitations}
\author{Giniyat~Khaliullin$^1$ and Satoshi~Okamoto$^2$}
\address{$^1$Max-Planck-Institut f\"ur Festkorperforschung, 
Heisenbergstrasse 1, D-70569 Stuttgart, Germany}
\address{$^2$The Institute of Physical and Chemical Research (RIKEN), 
Saitama 351-0198, Japan}
\date{\today}
\maketitle
\begin{abstract}
We investigate the collective behavior of orbital angular 
momentum in the spin ferromagnetic state of a
Mott insulator with $t_{2g}$ orbital degeneracy. 
The frustrated nature of the interactions leads 
to an infinite degeneracy of classical states. 
Quantum effects select four distinct orbital orderings. 
Two of them have a quadrupolar order, 
while the other states show in addition weak orbital magnetism. 
Specific predictions are made for neutron scattering 
experiments which might help to identify the orbital 
order in YTiO$_3$ and to detect 
the elementary orbital excitations. 
\end{abstract}
]
\narrowtext
\noindent

Recently, ``orbital physics'' has become an important topic in  
the physics of transition metal oxides\cite{TOK00,IMA98}.
Orbital degeneracy of low energy states 
and the extreme sensitivity of the magnetic bonds to the spatial
orientation of orbitals lead to 
a variety
of competing phases that are tunable by a moderate external fields. 
A well known example is ``colossal magnetoresistivity'' of manganites.
A ``standard'' view on orbital physics is based on pioneering 
works\cite{GOO55,KAN59},
developed and summarized in Ref.\cite{KUG82}.
In this picture, long-range coherence of orbitals 
sets in below a cooperative orbital/Jahn-Teller (JT) transition; 
spin interactions on every bond are then fixed
by the Goodenough-Kanamori rules.  
Implicit in this picture is that the orbital
splittings are large enough so that one can consider 
orbital populations 
as classical numbers. Such a classical treatment of orbitals is well
justified when orbital order is driven by strong cooperative lattice
distortion. 

However, interesting quantum effects are expected to come into play 
when orbital physics is dominated by electronic correlations.
Mott insulators with threefold $t_{2g}$ degeneracy (titanates, vanadates) 
are of particular interest in this respect, 
as the relative weakness of the JT coupling
for $t_{2g}$ orbitals and their large degeneracy 
enhance quantum effects. A picture of strongly fluctuating 
orbitals in titanate LaTiO$_3$
has recently been proposed on both experimental\cite{KEI00} 
and theoretical grounds \cite{KHA00,KHA01}.

Titanates are very interesting: they show a nearly continuous transition 
from an antiferromagnetic state in LaTiO$_3$ with unusually small moment to 
a saturated ferromagnetic one in YTiO$_3$ \cite {KAT97}. Yet another puzzle 
is the almost gapless spinwave spectrum of cubic symmetry 
recently observed in YTiO$_3$
by Ulrich {\it et al.} \cite {ULR02}. This finding is striking, 
because the orbital state predicted by band structure 
calculations \cite {MIZ96} is expected to break cubic symmetry 
of the spin couplings.
In principle, one may obtain isotropic spin exchange couplings by using
this orbital state \cite {ISH02};
however, this requires fine tuning of 
parameters, and rather small changes of the orbital state may reverse 
even the sign of the spin couplings \cite {ULR02}.

To gain more insight into the origin of the different 
behavior of orbitals in YTiO$_3$ and  
LaTiO$_3$, we ask in this Letter the following question: 
What kind of orbital state would optimize the 
superexchange (SE) energy {\it if we fix the spin state} 
to be ferromagnetic as in YTiO$_3$? 
We find a surprisingly rich physics: 
(i) there are several ``best'' orderings, some of them are 
highly unusual showing static orbital magnetism;
(ii) the resulting spin interactions 
respect cubic symmetry in accord with experimental data;
(iii) strong orbital fluctuations lead to 
anomalously large order parameter reduction.
These states have been overlooked previously, as their 
stabilization is an effect of quantum origin. 
We also discuss a mechanism which stabilizes ferromagnetic state in YTiO$_3$. 

{\it The} $t_{2g}$ {\it superexchange} can, in general, be written as 
$H_{SE} = J_{SE} \sum_{\langle ij \rangle} 
\bigl[ \bigl( {\vec S}_i \cdot {\vec S}_j +\frac{1}{4} \bigl)
\hat J_{ij}^{(\gamma)} ~ + ~\frac{1}{2}\hat K_{ij}^{(\gamma)} \bigl] $, 
where $J_{SE}=4t^2/U$ represents the overall energy scale.
The orbital operators $\hat J_{ij}^{(\gamma)}$ and $\hat
K_{ij}^{(\gamma)}$ depend on bond directions $\gamma(= a,b,c)$. 
Their explicit form in a $d^1(t_{2g})$ system like the titanates 
is given in Ref.\cite {KHA01}. 
If spins are all aligned 
ferromagnetically (i.e., ${\vec S}_i \cdot {\vec S}_j=1/4$), the above 
equation reduces to {\it the orbital Hamiltonian}:
\begin{equation}
H_{orb} = 
\frac{1}{2} \sum_{\langle ij \rangle} A_{ij}^{(\gamma)}. 
\label{eq:Horb}
\end{equation}
Hereafter, we use $r_1 J_{SE} $ as a unit of energy 
and drop a constant energy shift ($=-r_1 J_{SE}$). 
The coefficient $r_1=1/(1-3 \eta)$
originates from Hund's splitting of the excited $t_{2g}^2$ multiplet
via $\eta=J_H/U$. 
The orbital operators $A_{ij}^{(\gamma)}$
can be represented via a triplet ${\bf t}_i= (a,b,c)_i$ 
of constrained particles ($n_{ia}+n_{ib}+n_{ic}=1$) 
corresponding to $t_{2g}$ levels of $yz$, $xz$, $xy$ symmetry, 
respectively. Namely,
\begin{equation}
A_{ij}^{(c)}  =  n_{ia}n_{ja}+n_{ib}n_{jb} 
+ a_i^\dagger b_i b_j^\dagger a_j
+ b_i^\dagger a_i a_j^\dagger b_j, 
\label{eq:ABn_ab} 
\end{equation}
for the pair along the $c$ axis.
Similar expressions are obtained for the exchange 
bonds along the $a$ and $b$ axes by using in Eq.~(\ref{eq:ABn_ab}) 
$(b, c)$ and $(c, a)$ orbiton doublets, respectively.
The Hamiltonian (\ref{eq:Horb}) is the main focus
of this Letter; a theory of ferromagnetic magnons including 
spin-orbital effects will be presented in Ref.\cite {KHA02}.  

\begin{figure}[b]
\epsfxsize=0.8\columnwidth
\centerline{\epsffile{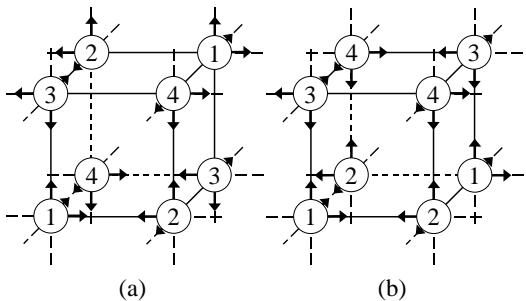}}
\caption{
Arrangement of the local quantization 
axes in states (a) and (b). Each state has four sublattices denoted 
by numbers.
Arrows show a direction of the quantization axes at each site.
They represent also a snapshot of local correlations of orbital 
angular momentum: on every bond, two out of three 
components $l_{\alpha}$ are correlated antiferromagnetically.} 
\label{fig:fig1}
\end{figure}

It is useful to look at the structure of $H_{orb}$ from different
points of view (see also Refs.\cite{KHA00,KHA01,KHA01a}): 
(i) On a given bond, the operator $A_{ij}^{(\gamma)}$
acts within a particular doublet of orbitals. 
Spinlike physics, that is, the formation of orbital singlets, 
is therefore possible. This is the origin of orbital fluctuations 
in titanates. (ii) On the other hand, interactions on different bonds
are competing: they involve different doublets, thus frustrating each other.
This brings about a Potts model like frustrations, from which 
the high degeneracy of classical orbital configurations follows. 
(iii) Finally, using 
the angular momentum operators 
$l_x= i (c^\dagger b- b^\dagger c)$ , 
$l_y= i (a^\dagger c- c^\dagger a)$ , 
$l_z= i (b^\dagger a- a^\dagger b)$  
of $t_{2g}$ level, 
one may represent the interaction as follows: 
\begin{eqnarray}
A_{ij}^{(c)} &=& (1-l_x^2)_i (1-l_x^2)_j + 
(1-l_y^2)_i (1-l_y^2)_j \nonumber \\
&&+ (l_x l_y)_i (l_y l_x)_j + (l_y l_x)_i (l_x l_y)_j ~.
\label{eq:ABn_l}
\end{eqnarray}
(For $b$ and $a$ bonds, the interactions are obtained by permutations of $l_z,l_y,l_x$.) 
We observe a pseudospin $l=1$ interaction 
of pure biquadratic form. Would $\vec l$ be a classical vector, it could
change its sign at any site independently. A local ``$Z_2$ symmetry'' and
the associated degeneracy of the classical states tells us that
angular momentum ordering, if any, is of pure quantum origin. 

The above points (i)$-$(iii) govern the underlying 
physics of the orbital Hamiltonian. 
By inspection of the global structure of $A_{ij}^{(\gamma)}$, 
one observes that 
the first two terms are definitely positive. However, the last two terms 
(which change the ``color'' of orbitals)
can be made negative on {\it all the bonds simultaneously}, if 
(i) on every bond, two particular components of 
$\vec l_i$ and $\vec l_j$ are antiparallel, and
(ii) remaining third components are parallel. 
For $c$ bonds the rule reads as: $l_{iz}l_{jz}$ and $l_{ix}l_{jx}$ are 
both negative, while $l_y$ components are parallel. 
(In terms of orbitons: $c_i$ and $c_j$ are 
in antiphase, $a_i$ and $a_j$ as well; but $b_i$ and $b_j$ have the 
same phase.) 
We find only two topologically different 
arrangements [called (a) and (b)], 
which can accommodate this curious mixture of ``2/3 antiferro'' plus
``1/3 ferro'' correlations (see Fig.1). 

For technical reasons, and also to simplify a physical picture, 
it is useful to introduce 
new quantization axes. This is done in two steps. 
First, we introduce local, sublattice specified quantization 
axes (see Fig.1):
\begin{eqnarray}
1 &:& (x,y,z) \rightarrow (x,y,z) , \hspace{2em}
2  :  (x,y,z) \rightarrow (-x,-y,z) , \nonumber \\
3 &:& (x,y,z) \rightarrow (-x,y,-z) , \hspace{0.5em}
4  :  (x,y,z) \rightarrow (x,-y,-z) . \nonumber
\end{eqnarray}
After corresponding sign transformations of 
$l_{i\alpha}$ and orbitons, 
one obtains a negative sign in front of the last two terms 
of $A_{ij}^{(\gamma)}$, 
thereby ``converting'' the interactions in a new axes 
to a fully ferromagnetic one. From now on, a sublattice structure
will not enter in the excitation spectrum.
From the above observations, it is also clear that all the components of $\vec l$ are equally 
needed to optimize all three directions. 
We anticipate, therefore, that the cubic diagonals are 
``easy'' (or ``hard'') axis for $\vec l$ fluctuations/orderings 
(recall that the Hamiltonian has no rotational symmetry). 
The second step is then to rotate quantization axes     
so that the new $z$ axis ($\tilde z$) is along 
one of the cubic diagonals (say, $[111]$). 
This is done as follows: $\vec l_i = \hat R ~\vec{\tilde l}_i$
(and ${\bf t} = \hat R ~\tilde {\bf t}$ for orbitons). 
The matrix 
$\hat R = \hat R_z(\pi/4) \hat R_y(\theta_0) 
\hat R_z(-\pi/4)$ with $\tan \theta_0 = \sqrt{2}$. 
$\hat R_\alpha(\theta)$ describes the rotation 
around $\alpha$ axis by angle $\theta$. 

By the above transformations, the orbital Hamiltonian obtains
well structured form and can be divided into two parts as 
$\tilde H_{orb} = H_{Ising} + H_{trans}$. 
Here, $H_{Ising}$ represents a ``longitudinal'' 
part of the interaction:  
\begin{eqnarray}
H_{Ising} = \frac{1}{6} \sum_{\langle ij \rangle}
\Bigl[\tilde l_{iz}^2 + \tilde l_{jz}^2 - 
\frac{3}{2} \tilde l_{iz}^2 \tilde l_{jz}^2 - 
\frac{1}{2} \tilde l_{iz} \tilde l_{jz} \Bigr]. 
\label{eq:H_ising}
\end{eqnarray}
$H_{trans}$ is responsible for fluctuations of 
transverse components $\tilde l_x$, $\tilde l_y$ and quadrupole 
moments of various symmetry. Its explicit, somewhat lengthy form
will be presented in Ref.\cite {KHA02}.  
In terms of $\tilde a$, $\tilde b$, $\tilde c$ orbitons, 
$H_{Ising}$ promotes 
a condensation of an appropriate orbiton, 
while $H_{trans}$ gives a dispersion of 
orbitons and their interactions. 

{\it Order and excitations}.---Now we are ready to discuss 
possible orbital orderings. Equation (\ref{eq:H_ising}) makes 
it explicit that there are two deep classical minima 
corresponding to two different ``ferro-type'' states: 
(I) quadrupole ordering (classically, $l_{iz}=0$) and
(II) magnetic ordering ($l_{iz}=1$, classically). 
We notice that the last, magnetic term in Eq.~(\ref{eq:H_ising}) is generated by 
quantum commutation rules when we rotate $H_{orb}$; 
this makes explicit that the $Z_2$ symmetry is only a classical one 
and emphasizes the quantum origin of orbital magnetism. 
We see below that states I and II are degenerate even on a quantum level. 
Noticing that an arbitrary cubic diagonal could be taken as $\tilde z$ 
and having in mind also two structures in Fig.1, one obtains a multitude of degenerate states. 
This makes, in fact, all the orderings very fragile. 

To be more specific about the last statement, we calculate the excitation
spectrum. Introduce first a quadrupole order parameter 
$Q = \langle(\tilde l_x^2+\tilde l_y^2)/2 - 
\tilde l_z^2\rangle  
=\langle n_{\tilde c}-
(n_{\tilde a}+n_{\tilde b})/2\rangle$,
and the magnetic moment $m_l = 
\langle\tilde l_z\rangle = 
i\langle\tilde b^\dagger \tilde a- \tilde a^\dagger \tilde b\rangle$
(we set $\mu_B =1$). 
State I corresponds to the condensation of the orbital $\tilde c$
(an equal mixture of the original $a,b,c$ states), 
while the ordering orbital in the state II is an imaginary one 
$(\tilde a - i\tilde b)/\sqrt{2}$.
Let us call the remaining two orbitals by $\alpha$ and $\beta$, 
which are the excitations of the model. 
Specifically, 
$\alpha=\tilde a$, 
$\beta=\tilde b$ in state I, and 
$\alpha = [\tilde a + i\tilde b + (1 + i)\tilde c]/2$, 
$\beta = [\tilde b - i\tilde a - (1 - i)\tilde c]/2$
in state II. 
Note that latter doublets can be regarded as magnons of 
orbital origin, describing fluctuations of the orbital magnetism. 
Remarkably, in terms of these doublet excitations, the linearized
Hamiltonian in a momentum space has the same form for both phases.
Namely, 
\begin{eqnarray}
H_{OW} &=& \textstyle{\sum_{\vec k} \Bigl[
n_{\alpha \vec k} + n_{\beta \vec k}
+\Bigl\{ \frac{1}{2}  (\gamma_1 + \gamma_2) \alpha_{\vec k} \alpha_{-\vec k}}
\nonumber \\
&& \hspace{1.7em}
\textstyle{
+\frac{1}{2} (\gamma_1 - \gamma_2) \beta_{\vec k} \beta_{-\vec k} - \gamma_3 
\alpha_{\vec k} \beta_{-\vec k}
+ H. c. \Bigr\} \Bigr] ,}
\label{eq:How2}
\end{eqnarray}
where 
$\gamma_1 = (c_x + c_y + c_z)/3$, $\gamma_2 = \sqrt{3}(c_y - c_x)/6$, 
$\gamma_3 = (2c_z -c_x -c_y)/6$, 
and $c_{\alpha}=\cos k_{\alpha}$. 
These expressions refer to state I. 
For state II, one should just interchange $\gamma_2$ and $\gamma_3$. 
However this does not affect the spectrum of the elementary excitations,
which is given by 
$\omega_{\pm}(\vec k)
= \sqrt{1-(\gamma_1 \pm \kappa)^2}$, where 
$\kappa = \sqrt{\gamma_2^2 + \gamma_3^2}$. 

Using matrix Green's functions for the ($\alpha,\beta$) doublet, 
we may calculate various expectation values: 
(i) quantum energy gain: $E_0=-0.214$ for all the phases; 
(ii) number of bosons not in the condensate: 
$\langle n_\alpha + n_\beta \rangle =0.54$.  
Thus, the order parameters are strongly reduced by fluctuations.  
We obtain $Q = 0.19$ ($m_l = 0$, of course) in state I, and 
$m_l = 0.19$ accompanied with small quadrupole moment $Q = -0.095$ in state II. 
To visualize the orbital patterns, we show in Fig.~2 
the electron density calculated including quantum fluctuations.

\begin{figure}[h]
\epsfxsize=1\columnwidth
\centerline{\epsffile{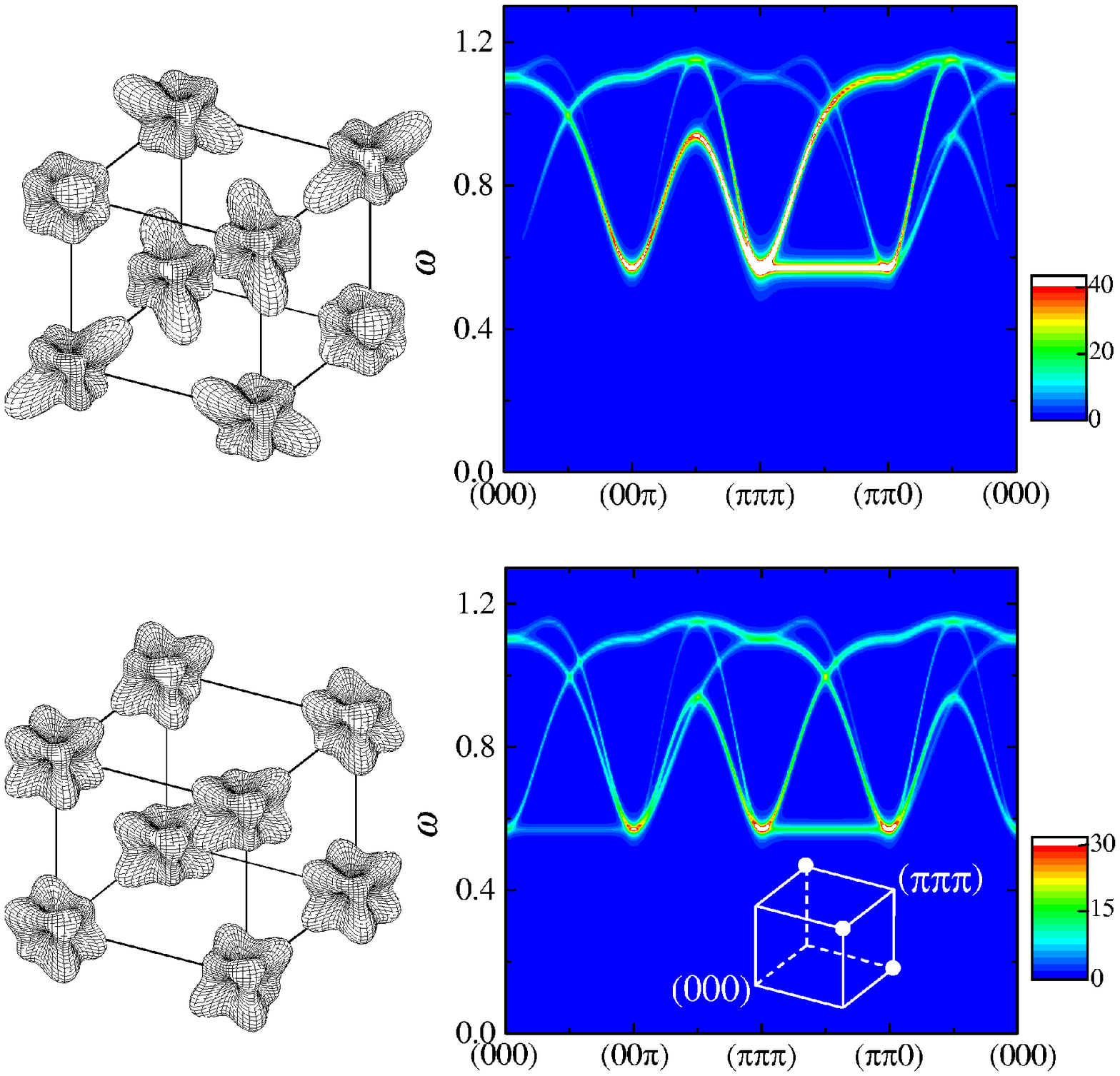}}
\caption{
$t_{2g}$ electron density (left) and 
intensity of the orbital contribution to the structure factor $S (\vec q, \omega)$ (right). 
Upper panel: Quadrupole ordered state I~(a). 
Because of the more spatial anisotropy of electron distribution,  
this state may be preferred by electron-lattice coupling. 
``Hot spot'' at ($\pi,\pi,\pi$) point is due to coherency factors
stemming from four sublattice structure. 
Lower panel: 
Magnetic state II~(a). 
Shown in the inset are the expected orbital Bragg peak positions.} 
\label{fig:fig2}
\end{figure}

The anomalous reduction of order parameters is due to the highly frustrated 
nature of the interactions in Eq.~(\ref{eq:Horb}). 
A special, nonspinlike feature of all orbital models is that orbitals are bond selective, 
resulting in a pathological degeneracy of classical states. 
This leads to soft modes 
[observe that $\omega_{\pm}(\vec k)$ is just flat along ($0,0,\pi$) and equivalent directions]; 
such soft modes were first noticed in Ref.\cite {FEI97}. 
Linear, ``spin-wave-like'' modes are, however, not true eigenstates even in the small momentum limit, 
as orbital models have no continuous symmetry. 
Therefore, an orbital gap must appear 
(see Ref.~\cite{KHA97} and, in particular, the discussion of the ``cubic'' model in Ref.~\cite{KHA01}). 
In the present model, we have estimated the orbital gap $\Delta = 0.57$ using a single mode 
approximation \cite {FEY72}.

Angular momentum fluctuations can directly be probed by neutron 
scattering measurements. 
We have calculated the structure factor
$S (\vec q, \omega) = \frac{1}{\pi} {\rm Im} \chi (\vec q, \omega)$, 
where  $\chi (\vec q, \omega)$ is the orbital angular momentum
susceptibility. We first 
obtain $\tilde \chi (\vec q, \omega)$ in a rotated basis. A noncollinear,
four sublattice orbital order generates 
in $S (\vec q, \omega)$ also an additional
terms $\tilde \chi (\vec q + \vec q_i, \omega)$ with shifted momenta.
For the arrangement (a) 
$\vec q_1= (\pi,   0, \pi), 
\vec q_2= (\pi, \pi,   0), 
\vec q_3= (  0, \pi, \pi)$, 
while for the structure (b), one has $\vec q_1= (\pi, \pi, \pi), 
\vec q_2= (\pi, \pi,   0), 
\vec q_3= (  0,   0, \pi)$. 
The results are shown in the right panels in Fig.~2. 
Neutron spectroscopy at higher than spin-wave energies
are necessary to detect ``orbital magnons.'' 
If one of the states II(a), (b) is realized, a static Bragg 
peak of orbital origin must show up. 
Reexamination of form-factor data
by Ichikawa {\it et al.} \cite {ICH00} using $t_{2g}$ spin densities
shown in Fig.2 is also desirable.

A remarkable feature of all the above phases is their robust property
to provide exactly the same spin couplings in all cubic directions, 
hence resolving the puzzling observation by Ulrich {\it et al.} \cite {ULR02}. 
A small spin-wave gap $\Delta_s$ in YTiO$_3$ further supports the theory. 
In state I~(a), we obtained a spin-orbit coupling $\lambda$-induced gap 
$\Delta_s \simeq A^2 / J_{SE}$, where $A \simeq \lambda^2/3 r_1 J_{SE}$. 
With $\lambda \sim 15-20$~meV and $J_{SE} \sim 40-50$~meV, 
one obtains $\Delta_s$ below 0.1~meV consistent with experiment. 
The gap is small due to high symmetry of the orbital ordering. 
On the other hand, a large classical gap $\Delta_s \sim A$ is obtained in states I~(b), II~(a), (b). 
Thus, the data of Ref.~\cite{ULR02} supports state I~(a) in YTiO$_3$. 

We discuss now our findings in a context of 
the interplay between different spin states in titanates.  
It is noticed that the quantum energy gain  
$E_F=-0.214 r_1 J_{SE}$
obtained above is only slightly
less than the energy of a spin AF/orbital disordered 
phase, $E_{AF}=-0.33 \frac{1-2\eta}{1-\eta} r_1 J_{SE}$ \cite {NOTE1}
stemming from a composite spin-orbital resonance. 
We argue now that the distortion of Ti-O-Ti bond angle $\theta$ due to a 
small size of Y ion further reduces the difference 
$\Delta E=E_{AF}-E_F$ and stabilizes the ferromagnetism of YTiO$_3$.  
This distortion (i) reduces hopping $t = t_0 \cos \theta$ \cite{KAT97}; hence, 
$J_{SE}=J_{SE}^{(0)} \cos^2 \theta$, 
and (ii) creates finite transfer $t' =t'_0 \sin \theta$ 
between $t_{2g}$ and $e_g$ orbitals [see inset of Fig.~3~(a)]. 
Here $t_0'/t_0 = t_{dp \sigma}/t_{dp \pi} \simeq 2$, 
and $t_{dp \sigma (\pi)}$ is the $\sigma (\pi)$ bond transfer 
between Ti $3d$ and O $2p$ orbitals.
The channel (ii) generates {\it unfrustrated ferromagnetic} interaction
(caused by Hund's coupling between $t_{2g}$ and $e_g$ electrons in
the excited state) with $J'=- \frac{4}{3} J_{SE}^{(0)} 
(U/\widetilde U)^2 \eta \sin^2 \theta$. 
Here $\widetilde U=U+\Delta_{cr}$ with $\Delta_{cr}$ 
being a cubic crystal field splitting. As a result, we obtain
$\Delta E =[-0.11 \cos^2 \theta 
+ 2 (U/\widetilde U)^2 \eta \sin^2 \theta ] J_{SE}^{(0)}$. 
As shown in Fig.~3~(a), $\Delta E$ reduces with decreasing $\theta$ and 
a transition from AF to ferrostate occurs 
at critical angle $\theta_c^{(0)}=136^\circ$. 
A ferromagnetic state with orbital order is expected to be further
supported by small JT energy gain.
By introducing $\delta E_{JT} = -0.04 J_{SE}^{(0)} \sim -3$~meV, 
we obtain $\theta_c=146^\circ$ 
as observed in titanates \cite{KAT97}. We notice that concominant
JT distortions associated with such a small $\delta E_{JT}$ are 
expected to be very weak.

Finally, it is stressed that orbital orderings are found to be weak; 
hence, large scale orbital fluctuations must be present. 
This implies strong modulations of the spin couplings, 
both in amplitude and sign, 
suggesting a picture of ``fluctuating exchange bonds'', where 
the magnetic transition temperature reflects 
only a time average of the spin couplings. 
The intrinsic quantum behavior of $t_{2g}$ orbitals in titanates 
may lead to a scenario, outlined in Fig.~3~(b), 
for an explanation of the puzzling 
``smooth'' connection between ferromagnetism 
and its antipode, the N\'eel state. This proposal can
be tested experimentally: 
we expect almost continuous magnetic transition under 
pressure or magnetic field that 
conserves the cubic symmetry of spin exchange constants. 

\begin{figure}
\epsfxsize=1\columnwidth
\centerline{\epsffile{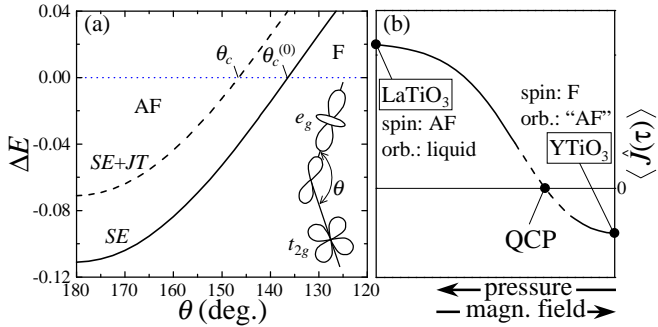}}
\caption{
(a) Energy difference (in units of $J_{SE}^{(0)}$) 
between AF and ferromagnetic states 
as a function of $\theta$. 
The solid line is obtained within the SE model.
Parameters are $\eta=0.12$ and $U/\Delta_{cr}=2.5$. 
The broken line includes small JT energy gain in the 
orbitally ordered ferromagnetic state. 
(b) Qualitative picture of the evolution of interactions in 
the spin channel. An average, static component of the spin exchange
``constant'' seen by coherent spin-waves is represented by a solid curve. 
Its sign depends on local correlations of orbitals. 
To the right of the critical point (QCP) these correlations are 
more antiferromagnetic, supported by noncollinear orbital orderings. 
To the left, the genuine ground state of the $t_{2g}$ system on  cubic lattice, 
an orbital disordered state supporting spin AF, is stabilized. 
In the proximity area, a fluctuating part of the overall SE interaction dominates, 
and separation of the spin and orbital degrees of freedom might no longer be possible.} 
\label{fig:fig3}
\end{figure}

To conclude, the ferromagnetic phase of the $t_{2g}$ superexchange system has several competing 
orbital orderings with distinct physical properties. 
The elementary excitations of these states are identified. 
We find that Ti-O-Ti bond distortion favors the ferromagnetic state. 
Theory well explains recent neutron scattering results in YTiO$_3$. 
An orbital driven quantum phase transition in titanates remains a major challenge for future study.

We thank B.~Keimer for stimulating discussions. 
Discussions with C.~Ulrich, P.~Horsch, R.~Zeyher, S.~Maekawa, J.~Akimitsu, 
and S.~Ishihara are acknowledged.


\begin{references}

\bibitem{TOK00} Y.~Tokura and N.~Nagaosa, Science {\bf 288},462 (2000).

\bibitem{IMA98}
M.~Imada, A.~Fujimori, and Y.~Tokura, Rev.\ Mod.\ Phys. {\bf 70}, 1039 (1998). 

\bibitem{GOO55} J.~B.~Goodenough, Phys.\ Rev.\ {\bf 100}, 564 (1955).

\bibitem{KAN59} J.~Kanamori, J.\ Phys.\ Chem.\ Solids {\bf 10}, 87
(1959).

\bibitem{KUG82}  K.~I.~Kugel and D.~I.~Khomskii, Sov.\ Phys.\ Usp.\
{\bf 25}, 231 (1982); Sov.\ Phys.\ Solid State\ {\bf 17}, 285 (1975).

\bibitem{KEI00} B.~Keimer {\it et al.}, Phys.\ Rev.\ Lett.\
{\bf 85}, 3946 (2000).

\bibitem{KHA00} G.~Khaliullin and S.~Maekawa, Phys.\ Rev.\ Lett.\
{\bf 85}, 3950 (2000).

\bibitem{KHA01} G.~Khaliullin, Phys.\ Rev.\ B 
{\bf 64}, 212405 (2001).

\bibitem{KAT97} T.~Katsufuji, Y.~Taguchi, and Y.~Tokura, 
Phys.\ Rev.\ B, {\bf 56}, 10145 (1997).

\bibitem{ULR02} C.~Ulrich {\it et al.}, preprint.

\bibitem{MIZ96} T.~Mizokawa and A.~Fujimori, Phys.\ Rev.\ B 
{\bf 54}, 5368 (1996); 
H.~Sawada and K.~Terakura, Phys.\ Rev.\ B {\bf 58}, 6831 (1998). 


\bibitem{ISH02} S.~Ishihara, T.~Hatakeyama, and S.~Maekawa, 
Phys.\ Rev.\ B {\bf 65}, 064442 (2002).  

\bibitem{KHA02} G.~Khaliullin and S.~Okamoto, (unpublished).

\bibitem{KHA01a} G.~Khaliullin, P.~Horsch, and A.~M.~Ole\'s, 
Phys.\ Rev.\ Lett.\ {\bf 86}, 3879 (2001).

\bibitem{FEI97}  L.~F.~Feiner, A.~M.~Ole\'s,  and J.~Zaanen,
Phys.\ Rev.\ Lett.\ {\bf 78}, 2799 (1997).

\bibitem{KHA97} G.~Khaliullin and V.~Oudovenko, Phys.\ Rev.\ B 
{\bf 56}, R14~243 (1997).

\bibitem{FEY72} R.~P.~Feynman, 
{\it Statistical Mechanics} (Benjamin, New York, 1972), Chap.~11.

\bibitem{ICH00}
H.~Ichikawa {\it et al.}, Physica B {\bf 281-282}, 482 (2000).

\bibitem{NOTE1}
The result of Ref.~[7] corrected for the finite values of $\eta$.

\end{references}
\end{document}